\documentstyle[11pt,Blois,epsfig]{article}

\def\be{\begin{equation}}
\def\ee{\end{equation}}
\def\bea{\begin{eqnarray}}
\def\eea{\end{eqnarray}}

\def\Pom{I\!\!P}

\begin{document}
\vspace*{2cm}
\begin{center}
\Large{\textbf{XIth International Conference on\\ Elastic and Diffractive Scattering\\
Ch\^{a}teau de Blois, France, May 15 - 20, 2005}}
\end{center}

\vspace*{2cm}
\title{INCLUSIVE DIFFRACTIVE DIS AT HERA}

\author{ MIKHAIL KAPISHIN (FOR THE H1 AND ZEUS COLLABORATIONS) }

\address{Joint Institute for Nuclear Research,\\
Joliot Curie 6, 141980 Dubna, Moscow region, Russia\\
E-mail: kapishin@sunse.jinr.ru}

\maketitle\abstracts{
Recent precision measurements of diffractive neutral current deep inelastic
{\it ep} scattering are performed by the H1 and ZEUS Collaborations at the HERA collider in a wide range 
of photon virtuality $Q^2$. The first measurements of the large rapidity gap cross sections in  charged 
current processes at high $Q^2$ are also presented. The results are compared with model predictions based 
on parton density functions obtained from a DGLAP QCD fit to the data.}

\section{Introduction}

\begin{figure}[!thb]
\hspace{35mm} \psfig{file=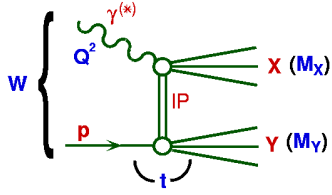,height=50mm}
\vspace{-2.3cm}
\caption{The diffractive process $\gamma^*p \rightarrow XY$ at HERA.}
\label{fig:diffhera}
\end{figure}

Diffractive processes in deep-inelastic $ep$ scattering (DIS) at HERA are characterized by the
presence of a large rapidity gap (LRG) between the leading proton (or the proton dissociation system $Y$) 
and 
the rest of the hadronic final state $X$ (Fig. \ref{fig:diffhera}). Here the variables $W$, $M_X$ and $M_Y$ 
denote the CMS energy of the virtual photon and the proton and the effective masses of the systems $X$ and 
$Y$, 
respectively. Within Regge phenomenology, diffractive processes are
described by the exchange of the Pomeron trajectory. As a result the Pomeron intercept can be  
extracted through a Regge fit to the data.
The presence of a hard scale - the photon virtuality $Q^2$ - enables
perturbative QCD to be applied to the data. Within the QCD framework the
diffractive events can be interpreted as processes in which a colour singlet
combination of partons is exchanged.

The structure of the colour singlet can be studied using a QCD approach based on the
hard scattering factorization theorem and parton density functions (PDFs) \cite{Collins}. 
In the charged current processes the structure of the color singlet could be probed by its 
coupling to the W boson. 

The diffractive reduced cross section is defined by

\begin{eqnarray}
\frac{d^4 \sigma^{ep \rightarrow eXp}}{dx_{\Pom} dt d\beta dQ^2} = \frac{4 \pi \alpha^2}
{\beta Q^4} \left( 1 - y + \frac{y^2}{2} \right) \cdot \sigma_r^{D(4)}(x_{\Pom},t,\beta,Q^2),
\end{eqnarray}

\noindent
where $y$ is the inelasticity, $t$ is the 4-momentum transfer
squared at the proton vertex, $x_{\Pom}$ is the longitudinal momentum fraction of
the incident proton carried be the colour singlet (${\Pom}$) and $\beta = x/x_{\Pom}$ is
the longitudinal momentum fraction of the colour singlet carried by the struck quark.
The $\beta$ variable for diffractive DIS processes is analogous to the Bjorken scaling
variable $x$ for inclusive DIS.  

$\sigma_r^D$ is related to the diffractive structure functions
$F_2^D$ and $F_L^D$ by:

\begin{equation}
\sigma_r^D = F_2^D - \frac{y^2}{1+(1-y)^2}F_L^D.
\end{equation}

\noindent
Thus $\sigma_r^D \simeq F_2^D$ is a good approximation except at the highest
$y$. The measurements are integrated over $|t|~<~1~{\rm GeV}^2$ because the leading
proton is not detected: $\sigma_r^{D(3)} = \int \sigma_r^{D(4)}dt$.

\section{Factorization properties of diffractive DIS}

The validity of the QCD hard scattering factorization for diffractive DIS was proven
by Collins \cite{Collins}. It states that at fixed $x_{\Pom}$ and $t$ the diffractive
cross section is a product of diffractive proton parton density functions 
$f_i^D$ and partonic hard scattering cross sections $\sigma^{\gamma^* i}$:

\begin{equation}
\sigma_r^{D(4)} \sim \sum \sigma^{\gamma^* i}(x,Q^2) \otimes f_i^D(x,Q^2,x_{\Pom},t).
\end{equation}

\noindent
$f_i^D$ are universal for diffractive $ep$ DIS processes (inclusive, dijet and 
charm production) and obey the DGLAP evolution equations. $\sigma^{\gamma^* i}$
are the same as for inclusive DIS. This approach allows us to test the diffractive exchange
within the perturbative QCD framework and extract diffractive parton density functions.
A NLO DGLAP QCD fit can be applied to diffractive DIS in analogy with inclusive DIS.

An additional assumption is made in the present analysis, that is, that
the $x_{\Pom}$ and $t$ dependences of diffractive parton densities
factorise from the $\beta=x/x_{\Pom}$ and $Q^2$ dependences (Regge factorisation) \cite{Ingelman}:

\begin{equation}
f_i^D(x_{\Pom},t,x,Q^2) = f_{\Pom}(x_{\Pom},t) \cdot f_i^{\Pom}(\beta,Q^2).
\end{equation}

\noindent
Here $f_{\Pom}$ is the Pomeron flux factor and $f_i^{\Pom}$ are Pomeron PDFs.
Although there is no firm proof in QCD for this assumption it is
approximately consistent with the present data.
The $x_{\Pom}$ dependence of the Pomeron flux factor was parameterized using a
Regge motivated form:

\begin{eqnarray}
f_{\Pom}(x_{\Pom}) = \int x_{\Pom}^{1-2 \alpha_{\Pom}(t)} e^{B_{\Pom}t} dt,    
\end{eqnarray}

\noindent
where $\alpha_{\Pom}(t) = \alpha_{\Pom}(0) + {\alpha'}_{\Pom}t$ is the Pomeron trajectory
with the intercept $\alpha_{\Pom}(0)$.

\section{Diffractive DIS cross sections}

\subsection{$W$ dependence of the diffractive cross section}

\begin{figure}[!thb]
\vspace{-0.3cm}
\hspace{45mm} \epsfig{figure=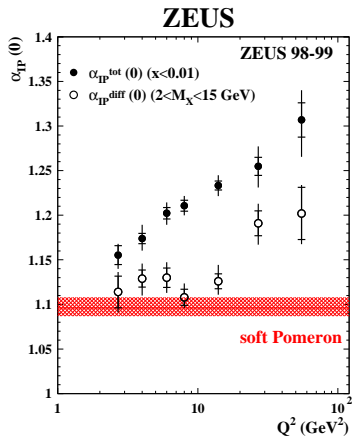,height=6cm}
\vspace{-0.4cm}
\caption{The Pomeron intercept as a function of $Q^2$ obtained from the $W$ dependence of the 
total $\gamma^*p$ cross section and the diffractive cross section $d\sigma/dM_X^2$.}
\label{fig:zeusalpha}
\end{figure}

ZEUS measured the $W$ dependence of the diffractive cross section $d\sigma/dM_X^2$ \cite{ZEUSFPC}. 
Using a new Forward Plug Calorimeter the acceptance was extended to higher $M_X$ and lower $W$ values in
comparison with the previous measurement \cite{ZEUSMx}. The results are presented in the kinematic range
$2.2 < Q^2 < 80~{\rm GeV}^2, 37 < W < 245~{\rm GeV}, M_X < 35$~GeV and $M_Y < 2.3$~GeV.
The Pomeron intercept $\alpha_{\Pom}$ was extracted using  a Regge motivated power-like fit 
$d\sigma/dM_X^2 \sim (W^2)^{2\alpha_{\Pom}(0)-2}$.
For $M_X < 2$ GeV the diffractive cross section shows a weak $W$ dependence and a strong 
decrease with $Q^2$ consistent with a higher 
twist behaviour. For larger $M_X$ values, a strong rise with $W$ and a weaker dependence on 
$Q^2$ are found. According to the optical theorem the total $\gamma^*p$ cross 
section is also characterized by a power-like dependence on $W$: $\sigma_{tot}^{\gamma^*p} \sim 
(W^2)^{\alpha_{\Pom}^{tot}(0)-1}$. The fit shows that the Pomeron intercept
extracted from the diffractive cross section is smaller than that from the total $\gamma^*p$ cross section 
(Fig. \ref{fig:zeusalpha}), but that the $W$ dependences of the cross sections  are similar: 
$2\alpha_{\Pom}-2 \approx \alpha_{\Pom}^{tot}-1$. The diffractive data are consistent with a "soft Pomeron" 
\cite{DL} at low $Q^2$, but also suggest the increase of the Pomeron intercept with $Q^2$, indicating possible Regge 
factorization breaking.  
 
\subsection{Diffractive reduced cross sections at medium and high $Q^2$}

\begin{figure}[!thb]
\hspace{35mm} \psfig{figure=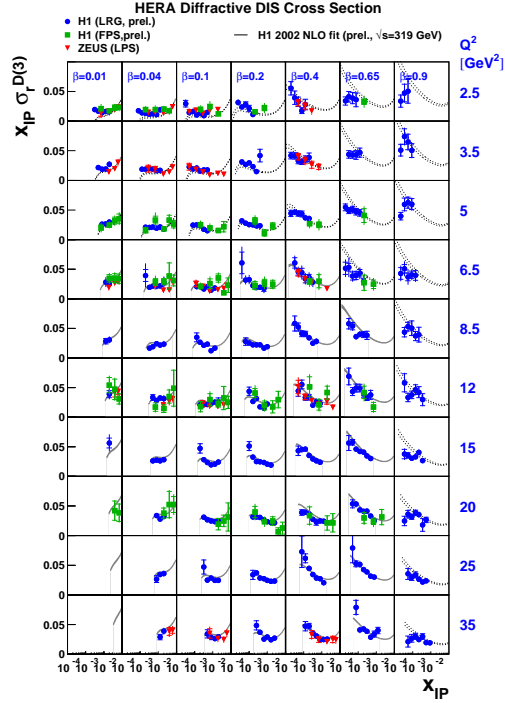,height=10cm}
\vspace{-1cm}
\caption{The diffractive reduced cross section measured in DIS processes with either 
a leading proton or a large rapidity gap in the final state.}
\label{fig:f2dall}
\end{figure}

Recent H1 and ZEUS measurements in which the leading proton is detected in proton spectrometers are 
described in \cite{Soares}. The selection of diffractive events containing a large rapidity gap (LRG) 
yields better statistical precision. In these events the leading proton is not detected and the kinematics are 
reconstructed from the hadronic system $X$. Figure \ref{fig:f2dall} shows a compilation of the H1 
and ZEUS measurements based on the selection of either a leading proton or a large rapidity gap in the 
final state. Good agreement between the two methods and the experiments points to a small contribution from 
proton dissociation processes to the LRG data.  

The H1 LRG data measured in the range
of $6.5 < Q^2 < 120 {\rm GeV}^2$ \cite{H1_2002} were used to 
perform a NLO QCD fit \cite{H1_QCD}.
In the fit the diffractive exchange is parameterized by light quark
singlet and gluon density functions at a starting scale $Q_0^2 = 3 {\rm GeV}^2$ which are evolved 
according to the NLO DGLAP equations. 
The $x_{\Pom}$ dependence is assumed to factorise from 
the $\beta$ and $Q^2$ dependences and is described by a phenomenological Regge flux factor (5).

The momentum fraction carried by gluons is estimated to be $75 \pm 15 \%$. 
This result is consistent with the result of the NLO QCD fit to the leading proton data performed recently 
by ZEUS \cite{ZEUSLPS}. PDFs extracted from the QCD fit to the diffractive cross sections can be 
used to test QCD factorization in the production of charm and dijets in $ep$ DIS at HERA and 
$p\bar{p}$ scattering at the TEVATRON\cite{Yamasaki}.

\begin{figure}[!thb]
\vspace{-0.2cm}
\hspace{45mm} \epsfig{figure=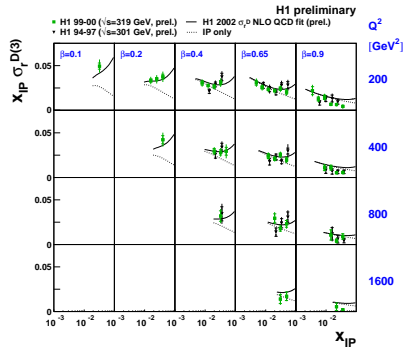,height=7.5cm}
\vspace{-2.8cm}
\caption{The diffractive reduced cross section as a function of $x_{\Pom}$ with the 
predictions based on the H1 NLO QCD fit.}
\label{fig:f2highq2}
\end{figure}

The H1 Collaboration has also measured the diffractive reduced cross section at high $Q^2$.
The results are presented in Fig.~\ref{fig:f2highq2}. The QCD prediction based on the 
PDFs 
extracted from the H1 NLO QCD fit
to the medium $Q^2$ data \cite{H1_2002,H1_QCD} gives a good description of the high $Q^2$ measurements, even 
though the NLO DGLAP evolution was performed over one order of magnitude in $Q^2$. The data show that
the contribution of a sub-leading trajectory is needed at high $x_{\Pom}$ and low $\beta$. 

\section{Charged current cross sections}

The charged current process
$ep \rightarrow \nu XY$ with a large rapidity gap in the final state has also been studied by H1 and ZEUS. 
In the diffractive CC process the W boson probes the partonic structure of the colour singlet exchange.
The LRG CC events were selected in the kinematic range: $Q^2>200 {\rm GeV}^2, y < 0.9, x_{\Pom} <0.05$. 

The LRG CC cross sections and their ratios to the inclusive CC cross section measured by H1 and ZEUS 
in the range $Q^2>200 {\rm GeV}^2, y < 0.9, x < 0.05$ are: \\

ZEUS:~$0.49 \pm 0.20({\rm st}) \pm 0.13 ({\rm sys})$ pb 

H1:~~~~~$0.42 \pm 0.13({\rm st}) \pm 0.09({\rm sys})$ pb \\
 
ZEUS:~$2.9 \pm 1.2({\rm st}) \pm 0.8({\rm sys}) \%$

H1:~~~~~$2.5 \pm 0.8({\rm st}) \pm 0.6({\rm sys}) \%$ \\
 
\noindent The results from the two experiments are in good agreement. 

The data were used to test predictions based on PDFs extracted from the diffractive NC 
DIS. A model based on PDFs extracted from the H1 LO QCD fit \cite{H1_1994} gives reasonable 
description of the ZEUS LRG CC data after statistical subtraction of the non-diffractive background.
The data are also consistent within the experimental errors with the 
non-diffractive distribution alone.

The H1 LRG CC cross sections are in good agreement with predictions of the recent H1 NLO QCD fit \cite{H1_QCD} 
assuming an additional contribution from a sub-leading reggeon trajectory.


\section*{Acknowledgments}

I thank my colleagues from H1 and ZEUS Collaborations 
who contributed to these results and gave assistance in preparing the talk.

\end{document}